\documentclass[11pt]{article}
\usepackage{amsmath}
\usepackage{amsfonts}
\usepackage{latexsym}
\usepackage[dvips]{graphicx}
\usepackage{epsf}

\allowdisplaybreaks

\newcommand{\be}{\begin{equation}}
\newcommand{\ee}{\end{equation}}
\newcommand{\bea}{\begin{eqnarray}}
\newcommand{\eea}{\end{eqnarray}}
 
\newcommand{\Zom}{\mathbb{Z}}
\newcommand{\Rom}{\mathbb{R}}
\newcommand{\Com}{\mathbb{C}}

\newcommand{\cC}{\mathcal{C}}

\newcommand{\cL}{\mathcal{L}}
\newcommand{\cO}{\mathcal{O}}

\newcommand{\x}{u}

\newcommand{\vb}{\bar{v}}

\newcommand{\I}{\mathrm{i}}
\newcommand{\e}{\mathrm{e}}
\newcommand{\p}{\partial}

\newcommand{\rmd}{{\rm d}}

\newcommand{\Li}{\mathrm{Li}_3}

\newcommand{\Fb}{\bar{F}}

\newcommand{\zb}{\bar{z}}

\newcommand{\vr}[1]{{\vec{r}\,}^{#1}}

\begin{document}

\begin{flushright} \small
 ITP--UU--07/21 \\ SPIN--07/14 
\end{flushright}
\bigskip

\begin{center}
 {\LARGE\bfseries Conifold singularities, resumming instantons  \\[1.2ex]
 and non-perturbative mirror symmetry }
\\[10mm]
 Frank Saueressig and Stefan Vandoren \\[3mm]
 {\small\slshape
 Institute for Theoretical Physics \emph{and} Spinoza Institute \\
 Utrecht University, 3508 TD Utrecht, The Netherlands \\
 {\upshape\ttfamily F.S.Saueressig,
 S.Vandoren@phys.uu.nl} }
\end{center}
\vspace{5mm}

\hrule\bigskip

\centerline{\bfseries Abstract} \medskip

We determine the instanton corrected hypermultiplet moduli space in type IIB
compactifications near a Calabi-Yau conifold point where the size of 
a two-cycle shrinks to zero. We show that D1-instantons resolve the 
conifold singularity caused by worldsheet instantons. Furthermore, by 
resumming the instanton series, we reproduce exactly the results obtained 
by Ooguri and Vafa on the type IIA side, where membrane instantons correct 
the hypermultiplet moduli space. Our calculations therefore establish 
that mirror symmetry holds non-perturbatively in the string coupling.

\bigskip

\hrule\bigskip
\section{Introduction}
\setcounter{equation}{0}

Recently, Calabi-Yau singularities have played a prominent role in ``bottom-up'' approaches connecting string theory to particle physics. The basic idea behind these constructions is to first locate a set of D-branes giving rise to the desired particle physics model at the singularity and later on perform the embedding into a compact Calabi-Yau threefold (CY$_3$). In this paper we study the simplest type of CY$_3$ singularities, the conifold, and determine the non-perturbative quantum corrections to the effective action for the (bulk) modulus which controls the size of the vanishing cycle. We expect that these type of corrections will become relevant when embedding the singularity into a full-fledged CY$_3$ compactification.

Geometrically, a conifold point is a point in the moduli space where the CY$_3$ becomes singular by developing a set of conical singularities (nodes) with base $S^2 \times S^3$. Locally these nodes can be resolved by either carrying out a deformation, by expanding the node into an $S^3$, or a small resolution expanding the node into an $S^2$. The process of shrinking a set of two-cycles $S^2$ to points and subsequently re-expanding the singularities into $S^3$'s (or vice versa) is called a conifold transition and connects moduli spaces of CY$_3$ with different Hodge numbers \cite{Candelas:1989ug,Aspinwall:1993nu}.

When approaching a conifold point by degenerating a complex structure ($S^3 \rightarrow 0$) and a K\"ahler structure ($S^2 \rightarrow 0$) in type IIB and type IIA string compactifications, respectively, the vector multiplet moduli space of the low energy effective action (LEEA) develops a logarithmic singularity. 
These singularities can be attributed to illegally integrating out D3 (D2) branes wrapping the vanishing cycles, which, at the conifold point, give rise to extra massless states \cite{SBH}. On the other hand we can also approach the conifold point by degenerating a K\"ahler structure on the type IIB or a complex structure on the type IIA side. This leads to a logarithmic singularity in the hypermultiplet sector of the LEEA. In this case, however, the theory has no BPS states which could wrap the vanishing cycles and could have an interpretation in terms of four-dimensional particles. One expects, however, that non-perturbative string effects originating from instantons \cite{Becker:1995kb} become important in this regime since the real part of their instanton actions are proportional to the volume of the shrinking cycle so that they are no longer suppressed in the limit where the cycle shrinks to zero. See fig. 1 for a schematic illustration. Indeed, as was shown in \cite{OV} in the context of IIA string theory, spacetime instanton effects survive in the effective field theory even in the rigid limit where gravity decouples, $M_{\rm Pl} \rightarrow \infty$.

More recently, new exact results were obtained in \cite{Robles-Llana:2006is} for IIB strings, in which the contributions coming from worldsheet instantons, D1-instantons and D(-1)-instantons to the effective action were determined. Since these results are obtained at a generic point in the moduli space, we can study the behavior near the conifold point, where a two-cycle shrinks to zero size and
gravity is decoupled. 
In this limit only worldsheet and D1-instanton corrections survive, and we obtain the resulting IIB hypermultiplet moduli space metric in the neighborhood of the conifold.

Our analysis allows us to perform a non-perturbative test of mirror symmetry, which states that the hypermultiplet moduli spaces in type IIA and type IIB on the mirror, after including all quantum corrections, must be the same. After resumming the instanton series on the IIB side, we determine the mirror map and show that the resulting hyperk\"ahler geometry is exactly the one obtained in \cite{OV}. This provides a nice demonstration of open string mirror symmetry on the hypermultiplet moduli space.

\begin{figure*}[t]
\setlength{\unitlength}{1cm}
\epsfxsize=1\textwidth
\begin{center}
\leavevmode
\epsffile{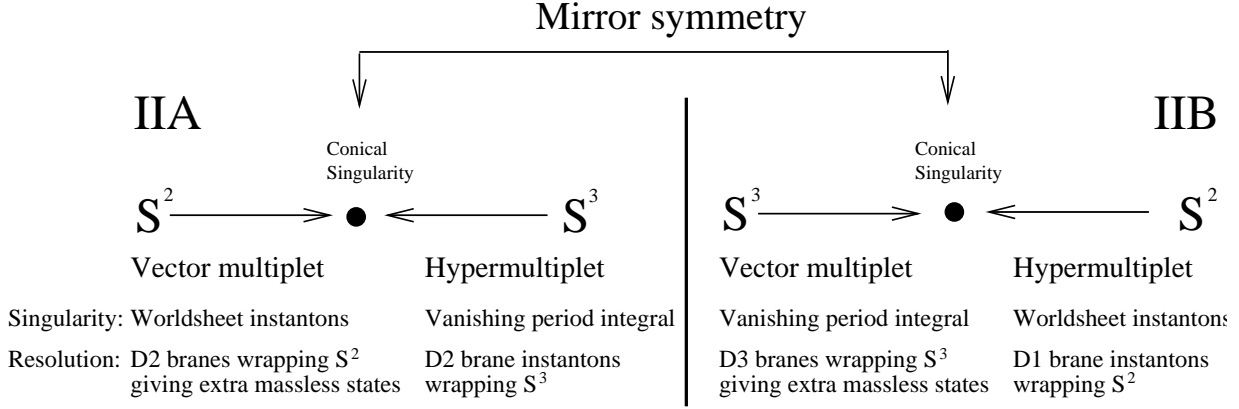}\,
\end{center}
\parbox[c]{\textwidth}{\caption{\label{fig}{\footnotesize Illustration of 
the CY$_3$ moduli space close to a conifold point. For more information on conifold singularities we refer to \cite{SBH,Greene:1996dh}.
}}}
\end{figure*}

\section{IIA: Summing up membrane instantons}
\setcounter{equation}{0}
%
In this section, we review the results of \cite{OV} who studied the
geometry of the type IIA hypermultiplet (HM) moduli space near a conifold 
singularity associated with vanishing three-cycle ${\cal C}_3$,
\begin{equation}
\mbox{IIA HM conifold limit:}\quad z = \int_{{\cal C}_3} \Omega 
\rightarrow 0\ .
\end{equation}
To decouple gravity, we consider the combined limit
\be\label{CFlimitIIA}
z\rightarrow 0 \, , \quad  \lambda \rightarrow 0 \, , \quad
{\rm with} \; \; \;   \frac{|z|}{\lambda} = {\rm finite} \, ,
\ee
where $\lambda$ is the string coupling constant. 
In this limit, the moduli space becomes a four-dimensional hyperk\"ahler 
space which at the classical level, develops a singularity at $z=0$. 
The metric can be written as
\begin{equation}\label{HK-metric}
{\rm d}s^2= \lambda^2 [V^{-1}({\rm d}t-\vec{A}\cdot {\rm d}{\vec y})^2
+V |{\rm d}{\vec y}|^2]\ .
\end{equation}
Here, $\vec {y}=(\x,z/\lambda,\bar z/\lambda)$ with $\lambda$ kept fixed and $\x$ and $t$ are the RR scalars originating from the expansion of the RR 3-form with respect to the harmonic three-forms associated with the vanishing cycle $\cC_3$ and its dual. The metric \eqref{HK-metric} is hyperk\"ahler if
\begin{equation}
V^{-1}\Delta V =0 \ ,\qquad \vec{\nabla} V = \vec{\nabla} \times
\vec{A}\ ,
\end{equation}
where 
\begin{equation}
\Delta = \partial_{\x}^2+4\lambda^2\partial_z\partial_{\bar z}\ .
\end{equation}
Classically, at string tree-level and for large $|z|$, the metric is determined by
\begin{equation}\label{cfsol}
V = \frac{1}{4\pi}\ln \Big(\frac{1}{z\bar{z}}\Big)\ , \qquad
A_\x = \frac{\I}{4 \pi} \ln\left( \frac{z}{\zb} \right) \, , \; A_z = 0 \, , \; A_{\zb} = 0 \, ,
\end{equation}
and has a logarithmic singularity. Using T-duality, which exchanges vector 
multiplets and hypermultiplets, this singularity has a counterpart
in the vector multiplet moduli space of the IIB theory compactified
on the {\it same} Calabi-Yau threefold, where it corresponds to the
appearance of massless black holes \cite{SBH}. In fact, the hyperk\"ahler
metric \eqref{HK-metric} is related to the vector multiplet moduli space
metric by the rigid c-map \cite{CFG,rcmap}. We demonstrate this in Appendix A. This 
will be important for us, since we use a similar mechanism for the mirror
theory in the next section.

In \cite{OV} Ooguri and Vafa studied the resolution of the singularity 
based on D2-brane instanton contributions. Thereby they focused on the situation where the period with respect to $\cC_3$ (A-cycle, say) vanishes while the dual period (from the B-cycle) remained finite. In this case membrane instantons wrapping the vanishing cycle 
generate exponential corrections to the hypermultiplet moduli space of the 
form $\exp(-|z|/\lambda)$ with $\theta$-angle $\exp(2 \pi \I \x)$, breaking the shift symmetry in $\x$ to a discrete subgroup. Membrane instantons wrapping the dual B-cycle decouple in the rigid limit \eqref{CFlimitIIA}, so that the shift-symmetry in $t$ is unbroken. The instanton corrected $V$ was then found to be \cite{OV}
\be\label{V-sum}
V = \frac{1}{4 \pi} \ln\left( \frac{\mu^2}{z \zb} \right) 
+ \frac{1}{2 \pi} \sum_{m \not = 0} K_0\left(2 \pi \frac{|m z|}{\lambda}
\right) 
\; {\rm e}^{ 2 \pi \I m \x }\ ,
\ee
for some constant $\mu$.
This instanton sum contains the zero-th order modified Bessel function, 
accompanied
by theta-angle-like terms set by the RR scalar $\x$. The Bessel function 
can further be expanded for large argument, yielding exponentially suppressed
terms of the form $\exp[-2\pi(|mz|/\lambda -im\x )]$ together with an infinite
power series in $\lambda$ that describe the perturbative fluctuations 
around the instantons.

To exhibit the resolution of the singularity, one can perform a 
Poisson resummation,
\begin{equation}
V=\frac{1}{4\pi} \sum_{n=-\infty}^{\infty}\left(\frac{1}{{\sqrt{(\x-n)^2
+z\bar{z}/\lambda^2}}}-\frac{1}{|n|}\right)+\rm{const}\ ,
\end{equation}
which leads to a regular metric at $z=0$.

In the case of $N$ three-cycles shrinking to zero size, it was argued in 
\cite{OV,Greene:1996dh} that this leads to hyperk\"ahler metrics with $\Com^2/Z_N$ singularities. This metric is again of the form \eqref{HK-metric}, with 
$V\rightarrow NV$. In the next sections, 
we will reproduce all these results from type IIB strings compactified on 
the mirror Calabi-Yau, in which $N$ now counts the number of vanishing 
two-cycles. Thereby, we perform a non-perturbative test of mirror symmetry.

\section{Conifold singularities in type IIA vector multiplets}
\setcounter{equation}{0}
%

To understand the origin of the conifold singularity in the IIB hypermultiplet
moduli space, it is insightful to first study its counterpart on the type 
IIA vector multiplet side. The two sectors are related by T-duality and, at string tree-level, the IIB hypermultiplet moduli space is obtained from the IIA 
vector multiplet moduli space by the c-map \cite{CFG,Ferrara:1989ik}. 
At a generic point in the moduli
space, the special geometry is determined by a holomorphic prepotential $F(X)$ homogeneous of degree two, which receives perturbative $\alpha^\prime$ corrections from the worldsheet conformal field theory and worldsheet instantons
\be
F(X) = F_{\rm cl}(X) + F_{\rm pt}(X) + F_{\rm ws}(X)\ .
\ee 
Here
\be\label{VMPP}
\begin{split}
F_{\rm cl}(X) = & \, \frac{1}{3!} \, \kappa_{abc} \, \frac{X^a X^b X^c}{X^1} \, , \qquad
F_{\rm pt}(X) =  \, \I \, \frac{ \zeta(3)}{2 (2 \pi)^3} \, \chi_E \, (X^1)^2 \, , \\
F_{\rm ws}(X) = & \,  - \I \,\frac{1}{(2 \pi)^3} \, (X^1)^2 \, \sum_{k_a} \, n_{k_a} \, {\rm Li}_3\left( {\rm e}^{2 \pi \I k_a X^a / X^1} \right) \, ,
\end{split}
\ee
with $\kappa_{abc}, \chi_E$ and $n_{k_a}$ the triple intersection numbers, 
Euler number and instanton numbers of the CY$_3$ \cite{Candelas:1990rm} respectively 
(see, e.g., \cite{Hori} for additional background).

Microscopically the scalar fields in the vector multiplet sector arise from expanding the K\"ahler form $J$ and the ten-dimensional NS two-form $\hat B$ in terms of harmonic two-forms $\omega_a$ of the CY$_3$ \cite{Bodner:1990zm},
\be
\hat B = B_2 + b^a \, \omega_a \; , \quad J = t^a \, \omega_a \; , \quad a = 2, \ldots , h^{(1,1)} + 1 \, .
\ee
These fields are combined into complexified K\"ahler moduli
\be
z^a = b^a + \I t^a \,  = \frac{X^a}{X^1} \, . 
\ee
It is useful to factor out $X^1$ and work with the holomorphic function $f(z)$
determined by
\begin{equation}
F(X) = (X^1)^2 f(z)\ ,
\end{equation}
which is also computed by the genus zero topological string amplitude.

We are now interested in the conifold limit of the prepotential given above. In the CY$_3$ geometry the conical singularity is obtained by shrinking the size of a holomorphic two-cycle $\cC_\star$ to zero:
\be\label{gcs}
\mbox{geometrical conifold singularity:} \quad t^\star \rightarrow 0 \, . 
\ee

We note, however, that the condition \eqref{gcs} is not sufficient for causing a singularity in the vector multiplet moduli space. Here the singularity arises if the {\it complexified} K\"ahler modulus is taken to zero:
\be\label{mscs}
\mbox{moduli space conifold singularity:} \quad z^\star \rightarrow 0 \, . 
\ee
This implies that we can avoid hitting the singularity by giving a non-vanishing real part $b^\star$ to the complexified K\"ahler modulus. Thus in the moduli space the conifold singularity is a line of complex codimension one\footnote{This is different from the five-dimensional case \cite{FrankPhD} where such singularities are of real codimension one, so that a generic trajectory moving on the moduli space will not be able to avoid the singularity.}.

We can now take the conifold limit \eqref{mscs} for the prepotentials \eqref{VMPP}. Henceforth we consider the case of a conical singularity where one particular complexified K\"ahler modulus $z^\star = k_a z^a$ (for one particular and fixed vector $k_a$) shrinks to zero\footnote{We will drop the ''$\star$'' in the following.}, while the others are frozen to constant values. By inspection one then finds that the second derivatives of $f$ (determining the metric) arising from $f_{\rm cl}(z)$ and $f_{\rm pt}(z)$ are regular in this limit. Applying the expansion formula \eqref{Li3exp} to the worldsheet instanton contribution, one obtains (we
denote $N=n_{k_a}$ for the fixed vector $k_a$)
\be\label{fcfl}
f_{\rm ws}(z) = \frac{N}{4 \pi \I} \, z^2 \, \ln(z) + \ldots \, ,
\ee
where the dots give rise to regular contributions in the Lagrangian. Computing
\be
\p_z f_{\rm ws} = \frac{N}{2 \pi \I} \, z \, \ln(z) + \ldots \, ,
\ee
one finds that this is in precise agreement with the singular behavior 
found in the IIB vector multiplet sector when 
going to the conifold point by shrinking $N$ 
Lagrangian three-cycles \cite{SBH}. In this case, however, $z$ is interpreted as a complex structure moduli arising from the periods of the holomorphic three-form of the CY$_3$. 

The qualitative results of this section have already been discussed by Strominger \cite{SBH}, where it was argued that the conifold singularities in the type IIA vector multiplet sector originate from  strong coupling effect involving worldsheet instantons.

\section{Conifold singularities in IIB hypermultiplets}
\setcounter{equation}{0}
\label{sect:5}

In this section, we derive the conifold singularities that arise in the
hypermultiplet moduli space of type IIB compactifications. As in 
section 2, this singularity can be obtained from the rigid c-map on
the vector multiplet sector of the IIA theory. Here, we will rederive it
in a different way, starting from a generic point in the (tree-level)  
hypermultiplet moduli space, and then taking the conifold limit in which 
gravity decouples. 
Our description of the hypermultiplet moduli space geometry uses the 
conformal tensor
calculus combined with methods used in projective superspace. 
In this way, $4n$-dimensional quaternion-K\"ahler geometry can be reformulated
in terms of $4(n+1)$-dimensional hyperk\"ahler geometry. For some 
background material we refer to \cite{deWit:1999fp,deWit:2001dj,
Bergshoeff:2004nf,Rocek:2005ij,RSV,deWit:2006gn}.

The tree-level hypermultiplet 
moduli space can be conveniently written down in projective 
superspace \cite{Gates:1984nk}, 
in terms of a contour integral representation \cite{Hitchin:1986ea} 
of the superspace Lagrangian density
\begin{equation}
\cL(v, \vb, x) = {\rm Im} \oint \, \frac{{\rm d} \zeta}{2 \pi \I \zeta} \, 
H(\eta^I(\zeta))\ ,
\end{equation}
in terms of $h_{1,2}+2$ $N=2$ tensor multipets
\begin{equation}\label{TM}
\eta^I(\zeta)=\frac{v^I}{\zeta}+x^I-{\bar v}^I\zeta\ ,
\end{equation}
consisting of $N=1$ real linear multiplets $x^I$ and $N=1$ chiral multiplets
$v^I$. The Lagrangian density satisfies
\begin{equation}
{\cal L}_{x^Ix^J}+{\cal L}_{v^I{\bar v}^J}=0\ ,
\end{equation}
and expresses the fact that the dual hypermultiplets parameterize a 
hyperk\"ahler manifold with $h_{1,2}+2$ commuting shift symmetries 
\cite{Hitchin:1986ea}. Tensor multiplets can be used because the 
hypermultiplet geometries 
we need to consider have enough commuting isometries\footnote{This is correct 
in the absence of three-brane and five-brane instantons, which are not relevant
for the purpose of this paper.}. The scalars of the tensor multiplets 
transform as a triplet under $SU(2)$ R-symmetry 
\be\label{SU2vec}
\vr{I} = \left[2 \, v^I, \, 2 \, \vb^I, \, x^I \right] \, , \quad \vr{I} \cdot \vr{J} = 2 v^I \vb^J + 2 v^J \vb^I + x^I x^J \, .
\ee
For a given prepotential $F(X)$ encoding the vector multiplet couplings,
the dual tensor multiplet Lagrangian after the (local) c-map can be 
obtained by evaluating the following contour integral \cite{Rocek:2005ij}
\be\label{cint}
\cL(v, \vb, x) = {\rm Im} \oint \, \frac{{\rm d} \zeta}{2 \pi \I \zeta} \, \frac{ F(\eta^\Lambda)}{\eta^0} \, .
\ee
Here $\eta^I = \{ \eta^0 , \eta^\Lambda \}$ with $\eta^0$ being the conformal compensator. The contour integral is taken around one of the roots $\zeta_+$ of $\zeta\eta^0$ and can be evaluated in a gauge invariant way \cite{Neitzke:2007ke}~\footnote{In \cite{sergei} a slightly more complicated formula for $\cL$ has been given, taking into account the logarithmic singularity at $\zeta = 0$. The two expressions, however, only differ by terms linear in $x^I$ and therefore lead to the same Lagrangian.}
\be\label{lagr-cmap}
\cL(v, \vb, x) =  - \frac{\I}{2 r^0} \left( F(\eta_+^\Lambda) - \bar{F}(\eta_-^\Lambda) \right) \\
=   - \frac{\I}{2 r^0} \left( (\eta_+^1)^2 f(z) - (\eta_-^1)^2 \bar{f}(\zb) \right) \,,
\ee
with
\be
\eta_+^\Lambda = \eta^\Lambda(\zeta_+) = x^\Lambda - \frac{x^\Lambda}{2} \left( \frac{v^\Lambda}{v^0} + \frac{\vb^\Lambda}{\vb^0} \right) - \frac{r^0}{2} \left( \frac{v^\Lambda}{v^0} - \frac{\vb^\Lambda}{\vb^0}\right) \, ,  
\ee
$\eta_-=(\eta_+)^*$ and $z^a = \eta_+^a / \eta_+^1$.

{}From the superspace Lagrangian density 
${\cal L}$, one can compute a tensor potential 
\cite{deWit:2006gn}
\begin{equation}
\chi (v,\vb,x)=-{\cal L}(v,\vb,x) + x^I {\cal L}_{x^I} ,
\end{equation}
where ${\cal L}_{x^I}$ denotes the derivative with respect to $x^I$. Dualizing
the tensors to scalars, this potential becomes the hyperk\"ahler potential
of the corresponding hyperk\"ahler cone above the quaternion-K\"ahler
manifold \cite{deWit:2001dj}. Therefore, this function
determines the entire low-energy effective action.
Using the homogeneity properties of ${\cal L}$, one can derive the 
identity 
\begin{equation}
\frac{1}{2} \left( \chi_{x^Ix^J} + \chi_{v^I \vb^J} \right) = 
{\cal L}_{x^Ix^J}\ .
\end{equation}
The components ${\cal L}_{x^Ix^J}$ then appear in the kinetic terms of the
scalars $x^I$ and $v^I$ in the effective Lagrangian.

Close to the conifold locus $f(z)$ is given by \eqref{fcfl}.
Substituting into \eqref{lagr-cmap} yields
\be\label{Lcfl}
\cL^{\rm cf}(v, \vb, x) 
= - \frac{N}{8 \pi r^0} \left( (\eta_+^1)^2 \, z^2 \, \ln(z) + (\eta_-^1)^2 \, \zb^2 \, \ln(\zb) \right) \,.
\ee
Here, only one tensor multiplet 
($v,\vb,x$ with $x=k_ax^a$ etc.) captures the degrees of 
freedom, and all others are frozen to constants.
As one can explicitly check, this function satisfies
\be\label{dgl1}
\left( \p_v \p_{\vb} + \p_x^2 \right) \cL^{\rm cf}(v, \vb, x) = 0 \, .
\ee
This is precisely the constraint coming from rigid $N=2$ supersymmetry and 
expresses the fact that the geometry is four-dimensional hyperk\"ahler. 
This is consistent
with the fact that in this limit, gravity is decoupled, and the target 
space of hypermultiplets becomes hyperk\"ahler\footnote{The rigid limit in 
special K\"ahler geometry was studied in detail in 
\cite{Billo:1998yr}. It would be desirable to have a similar study for 
hypermultiplets.}.

The function ${\cal L}$ is not yet to be compared with the function $V$
appearing in the hyperk\"ahler metric \eqref{HK-metric}. As shown in 
\cite{Hitchin:1986ea}, the relation is (in the dilatation gauge $r^0=1$)
\begin{equation}
V = r^0{\cal L}_{xx}\ .
\end{equation}
Straightforward computation shows that, up to an additive 
constant\footnote{This additive constant contributes to the parameter
$\mu$ in \eqref{V-sum} and depends on the particular CY$_3$ under 
consideration.},
\be\label{Lxxsing}
V = r^0{\cal L}_{xx} = - \frac{N}{4 \pi} \ln(z \zb) \, ,
\ee
which precisely matches \eqref{cfsol}. 
This shows that at string tree-level, mirror symmetry works.

\section{IIB: Resummation of D1-instantons}
\setcounter{equation}{0}
%
The starting point for including the D1-instantons is the modular invariant tensor potential \cite{Robles-Llana:2006is}
\be \label{TP}
\chi_{(1)} =  - \frac{r^0 \tau_2^{1/2}}{(2 \pi)^3}
\sum_{ k_a  } n_{k_a} \sideset{}{'} \sum_{m,n} \frac{\tau_2^{3/2}}{|m\tau + n|^3}\,
  \big( 1 + 2 \pi |m\tau + n|\, k_a t^a \big) \, \e^{-S_{m,n}}\ ,
\ee
with instanton action
\be
  S_{m,n} = 2\pi k_a \big( |m\tau + n|\, t^a - \I m\, c^a - \I n\, b^a
  \big)\ .
\ee
The primed sum is taken over all integers $(m,n) \in \Zom^2 \backslash (0,0)$.
Here we used the notation and conventions as in \cite{Robles-Llana:2006is}, and adapted the normalization in such a way that it is consistent with the prepotential \eqref{VMPP}. The formula \eqref{TP} contains all contributions coming from
both worldsheet instantons (sum over $n$) and D1-instantons (sum over $m$), which are the only relevant configurations that survive in the conifold limit\footnote{Reference \cite{Robles-Llana:2006is} also 
determined the contributions from
D(-1) instantons. They yield exponential corrections of the type
$\exp(-|m|\tau_2)$ and therefore vanish in the limit of vanishing string 
coupling. Similar arguments show that three-brane and five-brane instantons
decouple in the conifold limit \eqref{CFlimitIIB}.}. The instanton action contains the dilaton-axion complex
\begin{equation}
\tau = \tau_1 + \I \tau_2 = a + \I {\rm e}^{-\phi}\ ,
\end{equation}
and the string coupling constant is given by $\lambda = {\rm e}^\phi$.
Furthermore, the $c^a$ are RR scalars that generate the theta-angle like terms
for the D1-instantons. The relation between the ``microscopic'' scalars $\tau, z^a, c^a$ and the scalars appearing in the tensor multiplets \eqref{TM} is given by \cite{Berkovits:1998jh,Neitzke:2007ke,Robles-Llana:2006is}
\be
\tau = \frac{1}{(r^{0})^2} \left(\vr{0} \cdot \vr{1} + \I \, | \vr{0} \times \vr{1} | \right) \, , \;
z^a = \frac{\eta_+^a}{\eta_+^1} \, , \; c^a = \frac{(\vr{0} \times \vr{1}) \cdot (\vr{1} \times \vr{a})}{|\vr{0} \times \vr{1}|^2} \, .
\ee

Following the discussion in section 4 we compute the function $\cL_{xx}$ arising from \eqref{TP}. The corresponding calculation can be simplified by noting that both $\chi_{(1)}$ and $\cL_{xx}$ are invariant under local $SU(2)$ R-symmetry. Thus we can adopt a particular $SU(2)$ gauge, e.g., setting $x^0 = x^1 = 0, v^0 = \vb^0$ and then taking the derivatives of \eqref{TP} with respect to 
$x,v,\vb$. Re-expressing the result in gauge invariant variables we find
\be\label{LxxD1}
\cL_{xx}(x,v,\vb) = \frac{N}{4 \pi r^0} \sideset{}{'} \sum_{m,n} \frac{1}{|m \tau + n|} 
\, \e^{ -2 \pi \, \left( | m \tau + n| \, t - \I m c - \I n b \right) } \, .
\ee
To compare with the type IIA results obtained by Ooguri and Vafa we also have to take the conifold limit. On the type IIB side this corresponds to
\be\label{CFlimitIIB}
t \rightarrow 0 \, , \quad b \rightarrow 0 \, , \quad \tau_2 \rightarrow \infty \, , \quad {\rm with} \; \; \; \tau_2 |b + \I t| = {\rm finite} \, .
\ee
 Taking this limit requires resumming the instanton corrections appearing in \eqref{LxxD1}. For this purpose we split the double sum into the contributions coming from worldsheet instantons, $m=0$, and the D1-instantons plus their bound states, $m \not = 0, n \in \Zom$:
\be\label{Lxxsplit}
\begin{split}
\cL_{xx}(x,v,\vb) = & \, \frac{N}{4 \pi r^0}  \sum_{n \not = 0} \frac{1}{|n|} 
\e^{ -2 \pi \left( \, |n| \, t - \I n b \right) } 
+  \frac{N}{4 \pi r^0} \sum_{m \not = 0} \, \sum_{n \in \Zom} \frac{1}{|m \tau + n|} 
\, \e^{ -2 \pi \left( \, |m \tau + n| \, t - \I m c - \I n b \right) } \, .
\end{split}
\ee
The first term can be summed up easily
\be\label{int1}
\begin{split}
\frac{N}{4 \pi r^0} \, \sum_{n \not = 0} \frac{1}{|n|} \exp
\Big\{ -2 \pi \left( |n| t - \I n b \right) \Big\} 
& \, = - \frac{N}{4 \pi r^0} \ln\left(1 - {\rm e}^{2 \pi \I z} \right) + \mbox{c.c.} \\
& \, \simeq  - \frac{N}{4 \pi r^0} \ln\left( z \zb \right) \, ,
\end{split}
\ee
where we took the conifold limit $z = b+\I t \rightarrow 0$ in the second line. Observe that this expression precisely reproduces \eqref{Lxxsing}. In order to take the conifold limit in the second line of \eqref{Lxxsplit} we first carry out a Poisson resummation in $n$. Using the results of appendix \ref{SecB.2} we find
\be\label{int2}
\begin{split}
\frac{1}{4 \pi r^0} & \sum_{m \not = 0} \, \sum_{n \in \Zom} \frac{1}{|m \tau + n|} 
\, \e^{ -2 \pi \left( \, |m \tau + n| \, t - \I m c - \I n b \right) } \\
& = \frac{1}{2 \pi r^0} \sum_{m \not = 0} \, \sum_{n \in \Zom} 
K_0\left( 2 \pi |m \tau_2| \sqrt{t^2 + (b + n)^2} \right) \e^{2 \pi \I m (c - \tau_1 (b+n))} \\
& \simeq \frac{1}{2 \pi r^0} \sum_{m \not = 0} K_0(2 \pi \tau_2 |m z|) \; {\rm e}^{ 2 \pi \I m (c - \tau_1 b) } \, .
\end{split}
\ee
Here we have taken the conifold limit \eqref{CFlimitIIB} in the second step. Note that in this limit the sum over $n$ localizes such that only the $n=0$ part gives a non-zero contribution.

Combining \eqref{int1} and \eqref{int2}, we then obtain the D1-instanton
corrected $\cL_{xx}$ in the conifold limit
\be
N^{-1}V=r^0\cL_{xx} = \frac{1}{4 \pi } \ln\left( \frac{1}{z \zb} \right) 
+ \frac{1}{2 \pi } \sum_{m \not = 0} K_0(2 \pi \tau_2 |m z|) \; {\rm e}^{ 2 \pi \I m (c - \tau_1 b) }
\ee

Comparing this to the instanton corrected function $V$ on the type IIA side 
in \eqref{V-sum}, we find perfect agreement if we use the mirror map
\be\label{mirrormap}
\lambda = \tau_2^{-1} \,, \qquad z^{\rm IIA} = z^{\rm IIB} \,, \qquad \x = \pm(c - \tau_1 b) \, .
\ee
The second relation states that, under mirror symmetry, the complex structure modulus $z^{\rm IIA}$ is equated to the complexified K\"ahler modulus $z^{\rm IIB}$ associated with the vanishing cycles, while the relation between $u$ and $c - \tau_1 b$ is determined up to a sign only. Note that selecting the minus sign, eq. \eqref{mirrormap} is precisely the classical mirror map obtained in \cite{BGHL}. This shows that the classical mirror map does not receive quantum corrections 
once the conifold limit is taken. 

\medskip
\noindent
\textbf{Acknowledgments}

\noindent
We thank Cumrun Vafa for suggesting this project and for reading an earlier 
draft of this manuscript. We furthermore thank Bernard de Wit, 
Daniel Robles-Llana, Martin Ro\v{c}ek, Jan Stienstra and Ulrich Theis for 
valuable discussions. This work grew out of the 4th Simons Workshop in 
Physics and Mathematics. We thank the YITP and the Department of Mathematics 
at Stony Brook University for hospitality.\ FS is supported by the 
European Commission Marie Curie Fellowship no.\ MEIF-CT-2005-023966.
This work is partially supported by the European Union RTN network
MRTN-CT-2004-005104 and INTAS contract 03-51-6346.

\begin{appendix}
\section{Conifold singularities from the rigid c-map}
\setcounter{equation}{0}
The leading term in the type IIB vector multiplet prepotential can be determined by monodromy arguments
\be\label{VMPPIIB}
f_{\rm ws}(z) = \frac{1}{4 \pi \I} \, z^2 \, \ln(z) + \ldots \, ,
\ee
where $z$ is associated with the vanishing period of the holomorphic three-form of the CY$_3$.
We now interpret \eqref{VMPPIIB} as the prepotential underlying a rigid special K\"ahler geometry. We can then use the rigid c-map \cite{CFG,rcmap} to construct the dual hyperk\"ahler metric. For a general (rigid) prepotential $F(X^I)$ the resulting metric reads (up to a trivial rescaling)
\be\label{rcmet}
- \tfrac{1}{2} \rmd s^2 =  \I \left(\rmd F_I \, \rmd \bar X^I - \rmd \bar F_I \, \rmd X^I \right)
- N^{IJ} \left( \rmd B_I - F_{IK} \rmd A^K \right) \left( \rmd B_J - \Fb_{JL} \rmd A^L \right) \, , 
\ee
where $F_I = \p F / \p X^I$ and $N^{IJ}$ being the inverse of
\be
N_{IJ} = - \I \left( F_{IJ} -  \Fb_{IJ} \right) \, .
\ee
Evaluating \eqref{rcmet} for the prepotential \eqref{VMPPIIB} then yields
\be
\begin{split}
\rmd s^2 = & \, \frac{1}{2 \pi} \ln(z\zb) \, \rmd z \, \rmd \zb - \frac{\pi}{\ln(z\zb)} 
\left( \rmd B -  \frac{1}{2 \pi \I} \ln(z) \, \rmd A \right) \left(  \rmd B + \frac{1}{2 \pi \I} \ln(\zb) \, \rmd A \right) \, .  
\end{split}
\ee
Changing coordinates 
\be
B = 2 \lambda \, t \, , \; A = - 2 \lambda \, \x \, , 
\ee
one finds precisely the metric \eqref{HK-metric} obtained from the solution \eqref{cfsol}.
\section{Polylogarithms and resummation techniques}
\setcounter{equation}{0}
%
In this appendix we collect various facts used in the main part of the paper by giving a brief introduction to polylogarithmic functions and Poisson resummation in sections \ref{SecB.1} and \ref{SecB.2}, respectively.
\subsection{Polylogology}
\label{SecB.1}
We start by summarizing some properties of polylogarithmic functions, essentially following the appendix B of ref.\ \cite{LMZ}. For $0 < z < 1$, the k-th polylogarithm is defined via the series expansion
\be\label{PLD}
{\rm Li}_k (z) = \sum_{n=1}^{\infty} \frac{z^n}{n^k} \, .
\ee
It can be analytically continued to a multivalued function on the complex plane. Polylogarithms with different values of $k$ are related by
\be\label{polyrec}
z \frac{d}{dz} {\rm Li}_k(z) = {\rm Li}_{k-1}(z) \, .
\ee
For $k = 1$ we have
\be
{\rm Li}_1(z) = - \log(1-z) \, ,
\ee
which we used to sum up \eqref{int1}. From the definition \eqref{PLD} we find
\be
{\rm Li}_k(0) = 0 \; , \quad ( k \in \Zom) \quad {\rm and} \quad {\rm Li}_k(1) = \zeta(k) \, , \quad {\rm for} \quad k > 1 \, .
\ee
Polylogarithms at values $z$ and $1/z$ are related through the connection formula \cite{ConnectionFormula}, 
\be
{\rm Li}_k(z) + (-1)^k \, {\rm Li}_k(1/z) = - \frac{(2 \pi \I)^k}{k!} B_k\left( \frac{\log(z)}{2 \pi \I} \right) \, ,
\ee
where $B_k(\cdot)$ are the Bernoulli polynomials. For Li$_3(z)$ this yields
\be\label{CFLi3}
{\rm Li}_3(z) - {\rm Li}_3(1/z) = - \tfrac{1}{6} \log^3(z) - \tfrac{\I \pi}{2} \log^2(z) + \tfrac{\pi^2}{3} \log(z) \, .
\ee
{}From the point of view of the main part of the paper, it is more natural to work with the variable $x$, $z = {\rm e}^x$. In this case \eqref{CFLi3} becomes
\be\label{CFLix}
{\rm Li}_3(\e^x)  = {\rm Li}_3(\e^{-x}) - \tfrac{1}{6} x^3 - \tfrac{\I \pi}{2} x^2 + \tfrac{\pi^2}{3} x \, .
\ee
The conifold point corresponds to $x = 0$. At this point the function ${\rm Li}_3(\e^{-x})$ has a logarithmic branch point
\be\label{Li3expans}
{\rm Li}_3(\e^{-x}) \simeq q(x) \log(x) + p(x) \quad {\rm for} \; x \rightarrow 0 \, , 
\ee
where $q(x)$ and $p(x)$ are power series
\be
q(x) = \sum_{j=0}^\infty q_j x^j \; , \qquad p(x) = \sum_{j=0}^\infty p_j x^j \, .
\ee
Analytically continuing the ansatz \eqref{Li3expans} to Li$_3(\e^{x})$ using $\log(-x) = \log(x) + \I \pi$ and substituting into the connection formula \eqref{CFLix} we obtain the following expansion for small $x$
\be\label{Li3exp}
\Li \left(\e^{-x} \right) = - \frac{1}{2} x^2 \ln(x) + p(x)\ ,
\ee
where $p(x) = \zeta(3) - \zeta(2)x + \tfrac{3}{4}x^2 + \frac{1}{12} x^3 + \cO(x^4)$ is polynomial in $x$. With this identity it is then straightforward to determine the conifold limit of the prepotential \eqref{VMPP}.
\subsection{Poisson resummation}
\label{SecB.2}
Taking the conifold limit in the D1-brane instanton sector in 
section 5 requires a Poisson resummation in the worldsheet instanton number $n$. The technical details of this computation are collected in this appendix.

The basic ingredient for Poisson resummation is the following identity for the Dirac delta-distribution
\be
\sum_{n \in \Zom} \delta(y - na) = \frac{1}{a} \sum_{n \in \Zom} {\rm e}^{2 \pi \I n y/a} \, , \; a \in \Rom^+ \, .
\ee
Multiplying with an arbitrary function $f(x+y)$ and integrating over $ y \in \Rom$ gives the Poisson resummation formula
\be\label{PRF}
\sum_{n \in \Zom} f(x+na) = \frac{1}{a} \sum_{n \in \Zom} \tilde{f}(2 \pi n /a) \, \e^{2 \pi \I n x/a} \, . 
\ee
Here $f(x)$ and $\tilde f(k)$ are related by Fourier-transformation
\be
\tilde f(k) = \int_{-\infty}^{\infty} {\rm d}x \, f(x) \, \e^{-ikx} \; , \qquad f(x) = \frac{1}{2 \pi} \int_{-\infty}^{\infty} {\rm d}k \, \tilde f(k) \, \e^{ikx} \, . 
\ee
We now apply this resummation to the second term in \eqref{Lxxsplit}. Comparing
\be
\begin{split}
\sum_{n \in \Zom} \frac{1}{\left| m \tau + n \right|} \, \e^{- 2 \pi (| m \tau + n| \, t - \I n b)} 
= \sum_{n \in \Zom} \frac{1}{\sqrt{ (m \tau_2)^2 + (n + m \tau_1)^2 }} \, \e^{- 2 \pi ( \sqrt{ (m \tau_2)^2 + (n + m \tau_1)^2} \, t - \I n b)}
\end{split}
\ee
to the general formula \eqref{PRF} we identify
\be\label{B.15}
\tilde f(2 \pi n) = \frac{2 \pi}{(\alpha^2 + (2 \pi n + \gamma)^2)^{1/2}} \e^{- (\alpha^2 + (2 \pi n + \gamma)^2)^{1/2} t} \, , 
\ee
together with $a = 1, x = b, \alpha = 2 \pi m \tau_2$, and $\gamma = 2 \pi m \tau_1$. The (inverse) Fourier transform of \eqref{B.15} can be found using the following formula for Fourier cosine transformations \cite{IT}:
\be\label{FCT1}
\int_0^\infty {\rm d}x \, (x^2 + \alpha^2)^{-1/2} \, {\rm e}^{-\beta  \,(x^2 + \alpha^2)^{1/2}} \cos(xy)  =   K_0\left[\alpha (\beta^2 + y^2)^{1/2} \right] \, .
\ee
Substituting the result back into \eqref{PRF} then establishes the identity
\be\label{PR1}
\sum_{n \in \Zom} \frac{1}{\left| m \tau + n \right|} \, \e^{- 2 \pi ( | m \tau + n| t - \I n b)} = 2 \sum_{n \in \Zom} K_0\left(2 \pi |m \tau_2| (t^2 + (b+n)^2)^{1/2} \right) \e^{- 2 \pi \I m \tau_1 (b + n)} \, . 
\ee
This completes the derivation of the first step in \eqref{int2}.
\end{appendix}


\begin{thebibliography}{00}


\bibitem{Candelas:1989ug}
  P.~Candelas, P.~S.~Green and T.~H\"ubsch,
  {\it Rolling among Calabi-Yau vacua},
  Nucl.\ Phys.\  B {\bf 330} (1990) 49.



\bibitem{Aspinwall:1993nu}
  P.~S.~Aspinwall, B.~R.~Greene and D.~R.~Morrison,
  {\it Calabi-Yau moduli space, mirror manifolds and spacetime topology 
  change in string theory},
  Nucl.\ Phys.\  B {\bf 416} (1994) 414, 
  \texttt{hep-th/9309097}.

\bibitem{SBH}
  A.~Strominger,
  {\it Massless black holes and conifolds in string theory},
  Nucl.\ Phys.\  B {\bf 451} (1995) 96, {\tt hep-th/9504090};\\
B.~R.~Greene, D.~R.~Morrison and A.~Strominger,
  {\it Black hole condensation and the unification of string vacua},
  Nucl.\ Phys.\  B {\bf 451} (1995) 109, 
  \texttt{hep-th/9504145}.




\bibitem{Becker:1995kb}
  K.~Becker, M.~Becker and A.~Strominger,
  {\it Five-Branes, membranes and nonperturbative string theory},
  Nucl.\ Phys.\  B {\bf 456} (1995) 130,
  \texttt{hep-th/9507158}.


\bibitem{OV}
  H.~Ooguri and C.~Vafa,
  {\it Summing up D-instantons},
  Phys.\ Rev.\ Lett.\  {\bf 77} (1996) 3296,
  {\tt hep-th/9608079}.



\bibitem{Greene:1996dh}
  B.~R.~Greene, D.~R.~Morrison and C.~Vafa,
  {\it A geometric realization of confinement},
  Nucl.\ Phys.\  B {\bf 481} (1996) 513,
  \texttt{hep-th/9608039}.

\bibitem{Robles-Llana:2006is}
  D.~Robles-Llana, M.~Ro\v{c}ek, F.~Saueressig, U.~Theis and S.~Vandoren,
  {\it Nonperturbative corrections to 4D string theory effective actions from
  SL(2,Z) duality and supersymmetry},
  \texttt{hep-th/0612027}.

\bibitem{CFG}
  S.~Cecotti, S.~Ferrara and L.~Girardello,
  {\it Geometry of type II superstrings and the moduli of superconformal field theories},
  Int.\ J.\ Mod.\ Phys.\  A {\bf 4} (1989) 2475.

\bibitem{rcmap}
  J.~De Jaegher, B.~de Wit, B.~Kleijn and S.~Vandoren,
  {\it Special geometry in hypermultiplets},
  Nucl.\ Phys.\  B {\bf 514} (1998) 553,
  {\tt hep-th/9707262}.


\bibitem{Ferrara:1989ik}
  S.~Ferrara and S.~Sabharwal,
  {\it Quaternionic manifolds for Type II superstring vacua of Calabi-Yau
  spaces},''
  Nucl.\ Phys.\  B {\bf 332} (1990) 317.


\bibitem{Candelas:1990rm}
  P.~Candelas, X.~C.~De La Ossa, P.~S.~Green and L.~Parkes,
  {\it A pair of Calabi-Yau manifolds as an exactly soluble superconformal
  theory},
  Nucl.\ Phys.\  B {\bf 359} (1991) 21.



\bibitem{Hori}
K.~Hori, S.~Katz, A.~Klemm, R.~Pandharipande, R.~Thomas, C.~Vafa,
R.~Vakil and E.~Zaslow,
\emph{Mirror symmetry},
Clay mathematics monographs, AMS\@, Providence USA\@, 2003.

\bibitem{Bodner:1990zm}
  M.~Bodner, A.~C.~Cadavid and S.~Ferrara,
  {\it (2,2) vacuum configurations for type IIA superstrings: N=2 supergravity
  Lagrangians and algebraic geometry},
  Class.\ Quant.\ Grav.\  {\bf 8} (1991) 789.

\bibitem{FrankPhD}
  F.~Saueressig,
  {\it Topological phase transitions in Calabi-Yau compactifications of
  M-theory},
  Fortsch.\ Phys.\  {\bf 53} (2005) 5.


\bibitem{deWit:1999fp}
  B.~de Wit, B.~Kleijn and S.~Vandoren,
  {\it Superconformal hypermultiplets},
  Nucl.\ Phys.\  B {\bf 568} (2000) 475,
  \texttt{hep-th/9909228}.


\bibitem{deWit:2001dj}
  B.~de Wit, M.~Ro\v{c}ek and S.~Vandoren,
  {\it Hypermultiplets, hyperkaehler cones and quaternion-Kaehler geometry},
  JHEP {\bf 0102} (2001) 039, \texttt{hep-th/0101161}.


\bibitem{Bergshoeff:2004nf}
  E.~Bergshoeff, S.~Cucu, T.~de Wit, J.~Gheerardyn, S.~Vandoren and A.~Van Proeyen,
  {\it The map between conformal hypercomplex / hyper-Kaehler and
  quaternionic(-Kaehler) geometry},
  Commun.\ Math.\ Phys.\  {\bf 262} (2006) 411,
  \texttt{hep-th/0411209}.

\bibitem{Rocek:2005ij}
  M.~Ro\v{c}ek, C.~Vafa and S.~Vandoren,
  {\it Hypermultiplets and topological strings},
  JHEP {\bf 0602} (2006) 062, \texttt{hep-th/0512206}.


\bibitem{RSV}
  D.~Robles-Llana, F.~Saueressig and S.~Vandoren,
  {\it String loop corrected hypermultiplet moduli spaces},
  JHEP {\bf 0603} (2006) 081, {\tt hep-th/0602164}.

\bibitem{deWit:2006gn}
  B.~de Wit and F.~Saueressig,
  {\it Off-shell N = 2 tensor supermultiplets},
  JHEP {\bf 0609} (2006) 062, 
  \texttt{hep-th/0606148}.



\bibitem{Gates:1984nk}
  S.~J.~Gates, C.~M.~Hull and M.~Ro\v{c}ek,
  {\it Twisted multiplets and new supersymmetric nonlinear sigma models},
  Nucl.\ Phys.\  B {\bf 248} (1984) 157.


\bibitem{Hitchin:1986ea}
  N.~J.~Hitchin, A.~Karlhede, U.~Lindstr\"om and M.~Ro\v{c}ek,
  {\it Hyperkahler metrics and supersymmetry},
  Commun.\ Math.\ Phys.\  {\bf 108} (1987) 535.


\bibitem{Neitzke:2007ke}
  A.~Neitzke, B.~Pioline and S.~Vandoren,
  {\it Twistors and black holes},
  \texttt{hep-th/0701214}.


\bibitem{sergei}
  S.~Alexandrov,
  {\it Quantum covariant c-map}, {\tt hep-th/0702203}.

\bibitem{Billo:1998yr}
  M.~Billo, F.~Denef, P.~Fr\'e, I.~Pesando, W.~Troost, A.~Van Proeyen and D.~Zanon, {\it The rigid limit in special Kaehler geometry: From K3-fibrations to  special Riemann surfaces: A detailed case study},
  Class.\ Quant.\ Grav.\  {\bf 15} (1998) 2083,
  \texttt{hep-th/9803228}.





\bibitem{Berkovits:1998jh}
  N.~Berkovits,
  {\it Conformal compensators and manifest type IIB S-duality},
  Phys.\ Lett.\  B {\bf 423} (1998) 265,
  \texttt{hep-th/9801009}.

\bibitem{BGHL}
R.~B\"ohm, H.~G\"unther, C.~Herrmann and J.~Louis,
{\it Compactification of type IIB string theory on Calabi-Yau threefolds},
Nucl.\ Phys.\ \textbf{B569} (2000) 229,
{\tt hep-th/9908007}.



\bibitem{IT}
A.\ Erd\'elyi, et.\ al.\ , {\it Tables of integral transforms}, Vol.\ 1, McGraw-Hill Book Company, Inc., New York, 1954.





\bibitem{LMZ}
  J.~Louis, T.~Mohaupt and M.~Zagermann,
  {\it Effective actions near singularities},
  JHEP {\bf 0302} (2003) 053, {\tt hep-th/0301125}.


\bibitem{ConnectionFormula}
A.~P.~Prudnikov, Yu.~A.~Brychkov and O.~I.~Marikov, 
{\it Integrals and Series}, Vol.\ 3, Gordon and Breach, 1986.


\end{thebibliography}
\end{document}